\documentclass[aps,prl,twocolumn,superscriptaddress]{revtex4}

\usepackage{graphicx}
\usepackage{dcolumn}
\usepackage{bm}
\usepackage{times}
\usepackage{subfigure}
\usepackage{array}

\graphicspath{{./}{./figures/}}

\begin{document}


\title{Direct frequency comb spectroscopy of trapped ions}


\author{A.L. Wolf}\email{alwolf@few.vu.nl}\affiliation{Laser Centre Vrije Universiteit,
De Boelelaan 1081, 1081 HV Amsterdam, The Netherlands}
\affiliation{NMi van Swinden Laboratorium, Thijsseweg 11, 2629 JA Delft, The Netherlands}
\author{S.A. van den Berg}
\affiliation{NMi van Swinden Laboratorium, Thijsseweg 11, 2629 JA Delft, The Netherlands}
\author{W. Ubachs}\affiliation{Laser Centre Vrije Universiteit,
De Boelelaan 1081, 1081 HV Amsterdam, The Netherlands}
\author{K.S.E. Eikema}\affiliation{Laser Centre Vrije Universiteit,
De Boelelaan 1081, 1081 HV Amsterdam, The Netherlands}


\date{\today}

\begin{abstract}

Direct frequency comb spectroscopy of trapped ions is demonstated for the first time. It is shown that the \begin{math}4s\,^2\mathrm{S}_{1/2}-4p\,^2\mathrm{P}_{3/2}\end{math} transition in calcium ions can be excited directly with a frequency comb laser that is upconverted to 393 nm. Detection of the transition is performed using a shelving scheme to suppress background signal from non-resonant comb modes. The measured transition frequency of 
$f=761\, 905\, 012.7(0.5)$ MHz presents an improvement in accuracy of more than two orders of magnitude. \end{abstract}


\maketitle


Optical frequency comb lasers provide a phase coherent link between radio frequency sources and optical frequencies \cite{Diddams, Holzwarth}. As a result optical frequencies can be counted, which has made spectroscopy and optical clocks possible with an extremely high accuracy on the order of $10^{-16}-10^{-17}$ \cite{Ludlow, Rosenband}. In most experiments a continuous wave (CW) laser is used as a probe of an atomic or molecular transition, which is then calibrated against a frequency comb (see e.g. \cite{Fischer} for an experiment on atoms, \cite{Thompson} for references to measurements on tightly confined ions, and \cite{Wolf, Herrmann} for measurements on ions in the weak binding limit).\newline
\indent However, frequency combs can also be used for direct excitation, without the need of a CW laser. This is possible because the pulsed output of a comb laser (as seen in the time domain) is equivalent to many equidistant narrow band modes in the frequency domain. Direct frequency comb spectroscopy has been demonstrated in beam experiments \cite{Witte, Gerginov}, atomic vapour cells \cite{Fendel} and cold neutral atoms in magneto-optical traps \cite{Marian04, Fortier}, but not yet on cold ions in an ion trap. \newline
\indent Ion traps provide the opportunity to simultaneously trap different species. By laser cooling one type of ion, the other ions in the trap can be sympathetically cooled \cite{Larson, Hornekaer}. A system to trap and cool one ion species, can then be used without big modifications to trap and cool other types of atomic and molecular ions. Frequency combs offer a wide spectrum of frequencies through coherent broadening in nonlinear optical fibers \cite{Ranka} and doubling in nonlinear crystals or higher harmonic generation in gas jets \cite{Cavalieri}. The combination of direct frequency comb spectroscopy and ion trapping thus provides the possibility to perform spectroscopy on various ionic transitions in a single system.\newline

In this Letter we demonstrate for the first time direct frequency comb spectroscopy of ions in a trap. It is shown that this technique can be applied using an upconverted frequency comb at 394 nm, without amplification of the comb pulses. Furthermore, we combine it with a shelving scheme \cite{Bergquist} to suppress background signal from non-resonant comb modes, resulting in a very good signal to noise ratio despite the low power per comb mode. 
Calcium ions are used for this experiment because a more accurate calibration of the $4s\,^2\mathrm{S}_{1/2}-4p\,^2\mathrm{P}_{3/2}$ state is of interest for the search for a change of $\alpha$ over timespans of many billion years \cite{berengut}. Apart from this application, trapping and laser cooling of the calcium ion has been widely studied \cite{Urabe, Nagerl, Ritter}, in particular for atomic clocks \cite{Champenois, Matsubara} and quantum computation \cite{roos}. 



\begin{figure}
\includegraphics[width=85mm]{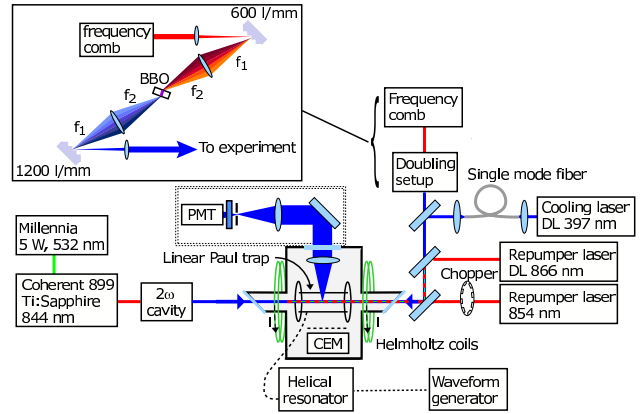}%
\caption{(Color online) Schematic layout of the laser system and linear Paul trap. DL=Diode laser, PMT=Photo-Multiplier Tube, 2$\omega$=frequency-doubling setup, CEM=Channel Electron Multiplier. Inset: Setup for chirped doubling of the frequency comb. Lenses are drawn for clarity, in practice only reflective optics are used. The comb spectrum is dispersed on a 600 l/mm grating before being doubled in a BBO-crystal. The colors in the doubled spectrum are overlapped again using a second, 1200 l/mm grating. f$_1= 20$ cm, f$_2=10$ cm.
 \label{fig:setup}}
\end{figure}

The spectroscopy is performed in a linear Paul trap, which is described in detail in \cite{Wolf}. 
The ion trap is mounted in a va\-cuum chamber, evacuated to a pressure of $3\times10^{-9}$ mbar. Calcium ions are produced by a two-step ionization process, involving the frequency comb laser. Evaporated calcium ions from an oven are first excited at 422 nm to the $4p\,^1$P$_1$ state by a frequency-doubled CW Ti:sapphire laser (Coherent 899), and subsequently ionized by the doubled frequency comb ($\lambda<391$ nm). This comb is also used for the spectroscopy on the ions (see below). The total system is schematically depicted in Fig. \ref{fig:setup}.

\begin{figure}
\includegraphics[width=80mm]{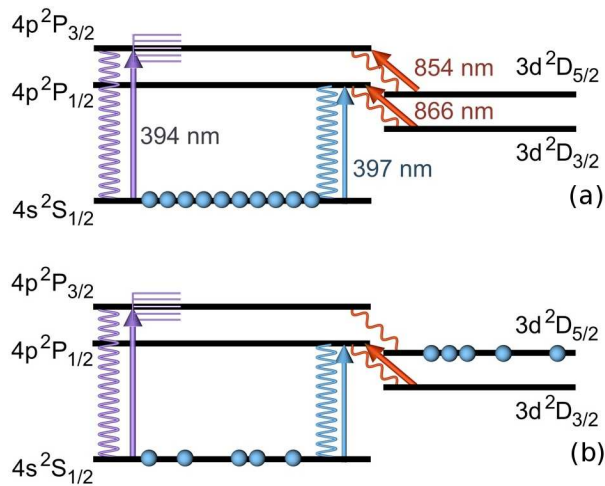}%
\caption{(Color online) Schematic view of the energy levels and the shelving principle. Arrows indicate laser beams, the wavy lines spontaneous fluorescence. The second repumper at 854 nm is periodically blocked using a chopper. If this repumper is present (situation (a)) all ions stay in the cooling cycle, a blocked repumper leads to transfer of the ions into the dark state, and the fluorescence from the cooling laser decreases (situation (b)).\label{fig:levels}}
\end{figure}

In order to reduce the Doppler effect on the single-photon spectroscopy transition, laser cooling is applied on the $4s\,^2\mathrm{S}_{1/2}-4p\,^2\mathrm{P}_{1/2}$ transition using a 397 nm diode laser (Toptica DL100). This laser is set at a detuning $\Delta f \approx 10$ MHz from resonance, locked to a wavemeter (Atos model LM-007) within $\approx 6$ MHz. The relevant energy levels are shown in Fig. \ref{fig:levels}. 
The spatial mode of this laser is cleaned up in a single mode fiber to reduce background from scattered light. The remaining 0.3 mW is focused to a beam diameter of $w_0=0.5$ mm in the trap. Since the excited ions have a 7\% chance to decay to the long-lived $3d\,^2\mathrm{D}_{3/2}$ state, a repumper diode laser at 866 nm is used ($P=1$ mW, Toptica DL100). Enough cooling power is available to crystallize the ion cloud, which reduces the Doppler broadening. The center of the trap is imaged onto a pinhole to remove background radiation and scattered light from the electrodes. Fluorescence from the $4s\,^2\mathrm{S}_{1/2}-4p\,^2\mathrm{P}_{1/2}$ transition is then observed using a photomultipler tube (PMT, Philips XP2020Q). From this fluorescence we determined a lifetime for the ions in the trap of $\tau \approx 8$ minutes.

Direct frequency comb spectroscopy relies on the the equivalence of phase coherent pulses (as emitted from a comb laser) and an equidistant mode structure in the frequency domain. These frequency comb modes can be described by the repetition frequency $f_{rep}$ of the pulses and a carrier-envelope offset frequency $f_{ceo}$. The frequency of the $n^{th}$ mode is equal to $f_n = \pm f_{ceo}+n\times f_{rep}$. Both $f_{rep}$ and $f_{ceo}$ are radio frequencies, which are locked to a frequency standard (in our case a Stanford PRS10 Rubidium atomic clock, referenced to the Global Positioning System).
 
The frequency comb laser for the experiment is based on Ti:sapphire. Chirped mirrors are used in the laser for dispersion management, and are chosen such that the output is maximized at the desired wavelength of $\lambda=788$ nm. This light is frequency doubled in a 3 mm BBO-crystal to obtain the right wavelength range to excite the $4s\,^2\mathrm{S}_{1/2}-4p\,^2\mathrm{P}_{3/2}$. 
By matching the angular dispersion induced by a grating to the wavelength derivative of the phase-matching angle \cite{Martinez} a wide bandwidth can be phase-matched. This method is used (Fig. \ref{fig:setup}, inset) to obtain a frequency comb in the blue with a FWHM of 13 nm and a power of P $\approx2$ mW.

The calcium ions are excited by focusing the upconverted comb light in the trap to an elliptical beam with a major axis ($1/e^2$ width) of 0.8 mm, and a minor axis of 0.4 mm. The generated spectrum is sufficiently broad for both probing the $4s\,^2\mathrm{S}_{1/2}-4p\,^2\mathrm{P}_{3/2}$ transition and ionizing excited neutral calcium. However, only one comb mode is resonant at the time in the ions because a single-photon transition is probed. All other $\approx 10^5$ modes do not contribute to the signal, but do give background signal due to scattered photons. This signal-to-noise issue is overcome by employing a `shelving' scheme (see e.g. \cite{Bergquist}). The scheme we use is depicted schematically in Fig. \ref{fig:levels}. Ions that are excited to the $4p\,^2\mathrm{P}_{3/2}$ state, have a 7\% chance to decay to the $3d\,^2\mathrm{D}_{5/2}$ state, which has a lifetime of $1.2$ s \cite{Gerritsma}. Trapped in this state they can no longer contribute to the cooling cycle. Since the fluorescence from the cooling laser is monitored, less signal will be observed in this situation (Fig. \ref{fig:levels}(b)). However, if a second repumper laser is used on the $3d\,^2\mathrm{D}_{5/2}$-$4p\,^2\mathrm{P}_{3/2}$ transition at $\lambda=854$ nm, ions will be pumped back into the ground state and participate in the cooling cycle (Fig. \ref{fig:levels}(a)).
 
\begin{figure}
\includegraphics[width=85mm]{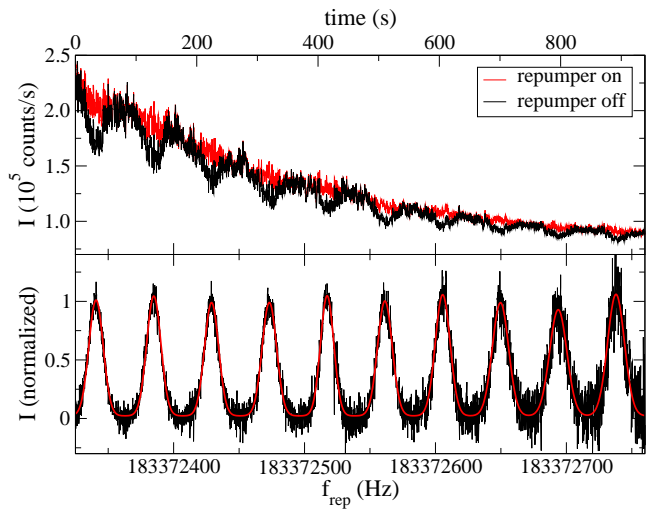}%
\caption{(Color online) Top: Measured fluorescence signal for the periods where the second repumper at 854 nm is on (red trace) and the periods where the repumper is off (black trace). The comb repetition frequency $f_{rep}$ is scanned against time (upper axis) and frequency (lower axis). Bottom: Normalized fluorescence signal (corrected for decay, thin black trace) and the corresponding fit (thick red trace).\label{fig:comb_scan}}
\end{figure}
In order to scan over the transition, the frequency comb repetition rate is varied. A small change in this parameter effectively leads to a frequency scan of the comb near the resonance. All shown results are measured by scanning in both increasing and decreasing frequency direction, in order to eliminate systematic effects due to calcium ion loss. The periods with and without the second repumper are alternated using a chopper at 100 Hz, and for each data point the fluorescence counts for both situations are recorded.  The fluorescence signal detected at 397 nm will now be unaffected by the presence of the comb laser for the periods where the 854 nm repumper is present, and only shows the loss of ions from the trap (`repumper on' in the upper part of Fig. \ref{fig:comb_scan}). The situation is different for the periods where the second repumper is blocked. In this case, every time a comb line comes into resonance with the $4s\,^2\mathrm{S}_{1/2}-4p\,^2\mathrm{P}_{3/2}$ transition, ions are pumped into the dark $3d\,^2\mathrm{D}_{5/2}$ state, so the measured fluorescence will decrease proportional to the excitation rate. Due to the periodic nature of the frequency comb, this signal is repeated for every comb mode that comes into resonance with the probed transition (`repumper off' in the upper part of Fig. \ref{fig:comb_scan}). The recorded signals for the situations with and without repumper are subtracted and corrected for the loss of calcium ions from the trap, resulting in a typical measurement curve as shown in the lower part of Fig. \ref{fig:comb_scan}. A comb of Gaussians is fitted to this curve where the width and height of each resonance is varied separately, but with fixed distances as given by the frequency comb spacing.

\begin{figure}
\includegraphics[width=85mm]{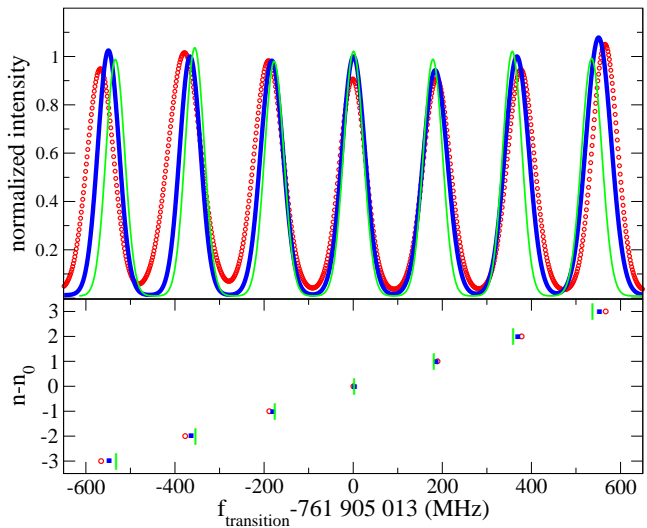}%
\caption{(Color online) Mode number determination: The upper part of the graph shows the fits to the scans at different frequency comb repetition frequencies, multiplied with a mode number $n_0$ to the approximate optical transition frequency (see text). Three different repetition frequencies ($f_{rep}$) are shown (frequency uncertainty margins are smaller than the shown data points): $f_{rep}=178$ MHz (thin green trace), $f_{rep}=183$ MHz (blue thick trace) and $f_{rep}=189$ MHz (circles, red trace). The lower part shows the peak centers deduced using mode numbers close to mode number $n_0$:  $f_{rep}=178$ MHz (green bars), $f_{rep}=183$ MHz (blue squares) and $f_{rep}=189$ MHz (red circles) . The overlap of the traces at $f_{trans}=761\,905\,012.7$ MHz can clearly be seen, and was checked for five different repetition frequencies (only 3 are shown for clarity).\label{fig:mode_number}}
\end{figure}

An important issue in frequency comb spectroscopy is the determination of the mode number $n_0$, usually based on previous measurements. The best known value for the $4s\,^2\mathrm{S}_{1/2}-4p\,^2\mathrm{P}_{3/2}$ transition has long been $\nu=25\,414.40 (15)$ cm$^{-1}$, presented in \cite{Edlen}, until a new value was reported of $\nu=25\,414.4137$ cm$^{-1}$ \cite{morton}. Because no uncertainty was known for this measurement, it has been re-evaluated recently leading to a best value of $\nu=25\,414.415 (2)$ cm$^{-1}$ \cite{litzen}. 
The resulting $1\sigma$ accuracy of 60 MHz is insufficient to assign the mode number with confidence.
We can, however, determine the mode by changing $f_{rep}$, which is graphically shown in Fig. \ref{fig:mode_number}. First an approximate mode number can be deduced from the earlier presented values, and the offset frequency $2\times f_{ceo}$ is taken into account (the factor two is due to the doubling of the comb). Using the selected mode number, one of the peak centers overlaps with the transition frequency, while the others deviate an integer times the repetition rate from this value. After the repetition frequency is changed by several MHz, the procedure is repeated. The peak centers of scans with different $f_{rep}$ will overlap at the peak with the correctly assigned mode number, but seperate on neighbouring peaks (Fig. \ref{fig:mode_number}).


\begin{figure}
\includegraphics[width=80mm]{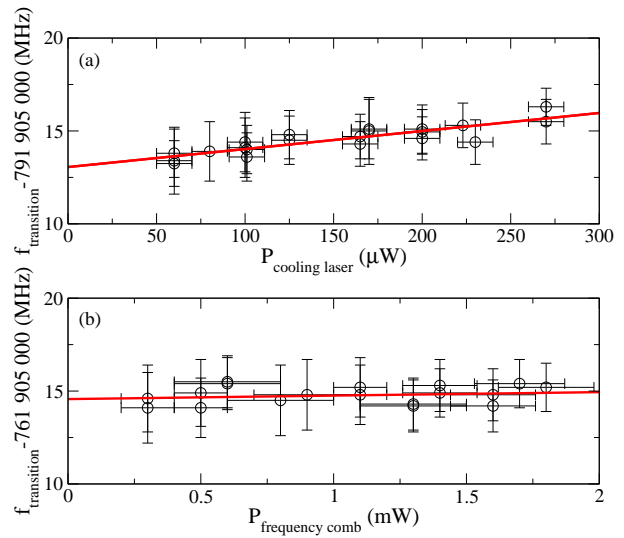}%
\caption{(Color online) (a) Measured data for light shifts induced by the cooling laser at $P_{comb}=1.9(0.2)$ mW  (black circles). (b) Measured data for light shifts induced by the frequency comb laser at $P_{cooling\ laser}= 200(10)$ $\mu$W. The weigthed fits to the data are also shown (red lines).\label{fig:stark}}
\end{figure}

We have investigated several systematic effects on the transition frequency. The strongest is due to
the always present 397 nm cooling laser on the  $4s\,^2\mathrm{S}_{1/2}-4p\,^2\mathrm{P}_{1/2}$ transition. Since its detuning is set at about half the natural linewidth, and the cooling laser couples to the same ground state as the transition that is measured, this laser can cause a significant AC Stark shift. The power dependence of this shift was measured (Fig. \ref{fig:stark} (a)), and fitted to give a frequency shift of $\Delta f_{trans}=9.7(1.3)$ kHz$/\mu$W$\times P_{cooling}$, where $P_{cooling}$ is the cooling laser power in $\mu$W. The corresponding zero-crossing is $f_{trans}=761\,905\,013.06 (0.21)$ MHz. In addition, at $P_{cooling}=200 (10) \mu$W the dependence of the measured transition frequency  on the frequency comb laser power was measured to be 
$\Delta f_{trans}=0.19(0.24)$ MHz$/$mW$ \times P_{comb}$, where $P_{comb}$ is the comb laser power in mW (Fig. \ref{fig:stark} (b)). The transition frequency is corrected for the corresponding shift at $P_{comb}=1.9$ mW of $\Delta f_{trans}=0.36 (0.46)$ MHz. The two repumper lasers do not cause a significant light shift, since the $866$ nm laser does not couple to either one of the levels involved in the measured transition, and the $854$ nm repumper is not present during the shelving intervals. The recoil shift is calculated to be 32 kHz, which is small compared to the measurement accuracy. Stark shifts due to the trapping fields are estimated to be below 4 kHz. Since the comb laser beam is linearly polarized, there is no first order Zeeman shift, while higher order Zeeman shifts are negligible compared to the measurement accuracy (see \cite{Wolf}). This leaves the statistical uncertainty due to the comb laser to consider. For a measurement time per data point of 0.1 s, the Allan deviation is estimated to be $10^{-9}$. Since a typical scan consists of about 3000 points measured over 10 minutes, this deviation averages down further and does not contribute significantly to the uncertainty budget. Once corrected for the measured shifts, the unperturbed transition frequency of the $4s\,^2\mathrm{S}_{1/2}-4p\,^2\mathrm{P}_{3/2}$ transition in $^{40}$Ca$^+$ is found to be $f=761\,905\,012.7(0.5)$ MHz ($1 \sigma$ uncertainty). This result is consistent with the  previous most accurate result of $f=761\,904\,994 (60)$ MHz \cite{litzen} \newline
\indent In conclusion, for the first time direct frequency comb spectroscopy on ions in a linear Paul trap has been demonstrated, using an upconverted comb laser and shelving detection. The obtained level of accuracy on the  $4s\,^2\mathrm{S}_{1/2}-4p\,^2\mathrm{P}_{3/2}$ transition in calcium ions is more than two orders of magnitude better than previous calibrations. The applicability of this method extends well beyond the probed ion and transition used for this experiment, if direct frequency comb spectroscopy were used on sympathetically cooled ions in a Paul trap.

\begin{acknowledgments}

The authors would like to thank U. Litz\'en for his updated calibrations of the $4s\,^2\mathrm{S}_{1/2}-4p\,^2\mathrm{P}_{3/2}$ transition, and S. Witte for discussions on the subject. 
This work is part of the Industrial Partnership Programme (IPP)
Metrology with Frequency Comb Laser (MFCL) of the Stichting voor
Fundamenteel Onderzoek der Materie (FOM), which is supported financially
by Nederlandse Organisatie voor Wetenschappelijk Onderzoek (NWO). The IPP
MFCL is co-financed by NMi, TNO and ASML.
\end{acknowledgments}

%




\end{document}